\documentclass[prl,aps,twocolumn,groupaddress]{revtex4-1}
\usepackage{latexsym}
\usepackage{graphicx}
\usepackage{verbatim}
\usepackage{multirow}
\usepackage{amsmath}
\newcommand{\beq}{\begin{eqnarray}}
\newcommand{\eeq}{\end{eqnarray}}
\usepackage{mathrsfs}
\usepackage{float}
\usepackage[usenames, dvipsnames]{color}
\usepackage{mathtools}
\usepackage{slashed}
\usepackage{physics}	% installed packages for typing maths and physics
\usepackage{graphicx}   % need for Fig.ures
\usepackage{epstopdf}
\usepackage{subfigure}  % use for side-by-side Fig.ures
\usepackage{hyperref}   % use for hypertext links, including those to external documents and URLs
\usepackage{bbold}
\usepackage{wasysym}
\usepackage{feynmp}
\usepackage[symbol]{footmisc}
\def \bs{\textbf}
\usepackage[latin1]{inputenc}
\usepackage{tikz}
\usetikzlibrary{decorations.pathmorphing}
\usetikzlibrary{shapes.misc}
\tikzset{cross/.style={cross out, draw=black, minimum size=8*(#1-\pgflinewidth), inner sep=0pt, outer sep=0pt},
cross/.default={1pt}}

\begin{document}

\title{Inequivalence of the zero-momentum Limits of Transverse and Longitudinal Dielectric Response  in the Cuprates}
\author{Chandan Setty}
\thanks{Corresponding author: csetty@illinois.edu}
\affiliation{Department of Physics, University of Illinois at Urbana-Champaign, Urbana, Illinois, USA}
\author{Bikash Padhi}
\affiliation{Department of Physics, University of Illinois at Urbana-Champaign, Urbana, Illinois, USA}
\author{Kridsanaphong Limtragool}
\affiliation{Department of Physics, University of Illinois at Urbana-Champaign, Urbana, Illinois, USA}
\author{Ali A. Husain}
\affiliation{Department of Physics, University of Illinois at Urbana-Champaign, Urbana, Illinois, USA}
\author{Matteo Mitrano}
\affiliation{Department of Physics, University of Illinois at Urbana-Champaign, Urbana, Illinois, USA}
\author{Peter Abbamonte}
\affiliation{Department of Physics, University of Illinois at Urbana-Champaign, Urbana, Illinois, USA}
\author{Philip W. Phillips}
\affiliation{Department of Physics, University of Illinois at Urbana-Champaign, Urbana, Illinois, USA}

\begin{abstract}
We address the question of the mismatch between the zero momentum limits of the transverse and longitudinal dielectric functions for a fixed direction of the driving field observed in the cuprates. This question translates to whether or not the order in which the longitudinal and transverse momentum transfers are taken to zero commute. While the two limits commute for both isotropic and anisotropic Drude metals, %in the presence of weak impurity scattering,
we argue that a scaleless vertex interaction that depends solely on the angle between scattered electron momenta is sufficient to achieve non-commutativity of the two limits even for a system that is inherently isotropic. We demonstrate this claim for a simple case of the Drude conductivity modified by electron-boson interactions through appropriate vertex corrections, and outline possible consequences of our result to optical and electron energy loss spectroscopy (EELS) measurements close to zero momentum transfer.\end{abstract}

\maketitle
\textit{Introduction:} It is well known \cite{Nozieres1999} that in the presence of a time-dependent homogeneous external field, the transverse and longitudinal ($\bs q$ perpendicular and parallel to the driving field respectively) dielectric constants are equal in that 
\begin{equation}
\epsilon_{\perp}(\textbf q \rightarrow \bs 0, \omega) = \epsilon_{\parallel}(\textbf q \rightarrow \bs 0, \omega).
\label{equivalence}
\end{equation}
 Simply put, this equality states that since it is the relative direction of the driving field and the wave vector that determines if a response is longitudinal or transverse, any such distinction must disappear in the $\textbf q \rightarrow 0$ limit. 
 However, various reflection and transmission EELS measurements (probes of the longitudinal dielectric function) performed in the normal state of the copper oxide superconductors~\cite{Zhao1989, Muller1991,Ritter1990,Nakai1990,Ishida1999,Schulte2002, Abbamonte2017}, exhibit plasmon lineshapes at low momenta between 0.1-1eV that are considerably different from optical measurements using Fourier Transform Infrared spectroscopy (FTIR) and ellipsometry (probes of transverse dielectric function)~\cite{Timusk2005, Bozovic1990,Leggett2016, Koshizuka1993,vanderMarel2003, Hwang2007,Onellion1999, Presura2003}. For instance, near zero-momentum, the fractional power-law dependence of the transverse dielectric function is not observed in measurements of the longitudinal dielectric function. Motivated by this observation, we address the following question: Does the zero momentum equivalence between the longitudinal and transverse dielectric functions (Eq.~\ref{equivalence}) continue to be valid in the presence of strong interactions indicative of the strange metal phase of the cuprates?\par
Of course, one might be tempted to invoke the role of spatial asymmetry as the operative mechanism underlying the failure of Eq.~(\ref{equivalence}). In an isotropic system,  Eq.~\ref{equivalence} holds regardless of how the functions $\epsilon_{\perp}(\textbf q \rightarrow \bs 0, \omega)$ and $\epsilon_{\parallel}(\textbf q\rightarrow \bs 0, \omega)$ are evaluated, i.e, with a varying direction of $\bs q$ and fixed driving field, or vice-versa. In the presence of anisotropy, however, this distinction becomes important. Clearly, when the transverse and longitudinal dielectric functions are evaluated for different orientations of the driving field, %(with the direction of the momentum transfer fixed while taken to zero)
one can trivially expect  Eq.~\ref{equivalence} to fail since the driving field sees different electronic and/or lattice potentials along different directions. On the other hand, the scenario more relevant for a comparison between optical and EELS measurements occurs when the direction of the driving field is fixed along a certain axis. In this case, it is natural to expect that Eq.~\ref{equivalence} must strictly continue to hold just as in a system with rotational symmetry. \par
Indeed according to our expectations for a fixed driving field, Eq.~\ref{equivalence} continues to hold even in the presence of anisotropy. The reason for this is simple to understand$-$the frequency of the driving field and the scattering rate together set energy scales in the problem, and since all energy scales emanating from the external field coupled to the electrons are compared to these scales, it does not matter whether the $\textbf q$ vector is taken to zero along the direction of the driving field or perpendicular to it. Hence Eq.~\ref{equivalence} holds true for both isotropic and anisotropic simple metals as long as the driving field is held constant.  As a result, the answer to this experimental conundrum lies elsewhere.

In this Letter, we identify a class of interactions which renders the breakdown of Eq.~\ref{equivalence} for a fixed direction of the driving field, even if the system is  inherently isotropic. Since they appear through vertex corrections and depend solely on the relative angles between the momenta of the scattered electrons, the applied external field has the additional effect of picking a direction in real space but does not add any additional scale to the interactions. This feature of the interaction distinguishes whether the wave vector is taken to zero along the driving field or perpendicular to it. We demonstrate this claim for a simple case of the Drude conductivity modified by electron-boson interactions through appropriate vertex corrections.

In recent experimental literature, the agreement between optical methods~\cite{Presura2003} and EELS~\cite{Nakai1990} data close to zero-momentum has been argued in certain contexts~\cite{VDM2015, Leggett2016}. However, no theoretical study has systematically addressed the relationship between optics and EELS in the limit of zero momentum. At this point, an obvious set of practical considerations with regards to the interpretation and comparison of optics and EELS data could be raised. First, despite the oblique angle corrections that are made in typical optical ellipsometry experiments, there are contributions from both transverse and longitudinal dielectric functions to the transmittance.   Second, as a practical matter, EELS cannot achieve a true zero momentum limit as a result of limitations in momentum resolution.  Third, as reflection EELS probes a surface response, the surface-to-bulk correspondence of the density response may be nontrivial. Legitimate as these  considerations might be, the question of whether, \textit{in principle}, Eq.~\ref{equivalence} is valid for a fixed driving field still remains. \par
 \textit{Order of limits:} In experiments, finite momentum longitudinal (transverse) dielectric response measurements are performed by setting the transverse (longitudinal) wave vector $\textbf{q}_{\perp}$ ($q_{\parallel}$) transfer to zero or close to zero (after oblique angle corrections etc). Therefore, the zero momentum transfer in the longitudinal (transverse) case is obtained by taking $q_{\parallel}$ ($\textbf{q}_{\perp}$) to zero \textit{after} one sets the perpendicular (parallel) component to zero. At an operational level, this is equivalent to taking the limits of the transverse and longitudinal momenta to zero in different order. Our goal here will be to uncover the circumstances under which the commutator $\left[\lim_{q_{\parallel}\rightarrow 0}, \lim_{\textbf{q}_{\perp}\rightarrow \bs 0}\right]$ of limits  of the response function does not vanish. \par
 \textit{Drude case:} That the equality in Eq.~\ref{equivalence} holds for simple metals can be argued directly from the Mermin formula \cite{Mermin1970} for the susceptibility in the presence of weak impurity scattering. The longitudinal conductivity is related to the polarization function through the relation $\sigma_{\parallel}(\textbf q, \omega) = \frac{i \omega}{4 \pi} V_q \Pi(\textbf q, \omega)$, where $V_q = \frac{4\pi e^2}{q^2}$. The Mermin susceptibility is given by $\chi_M(\vec q, \omega) = \frac{\chi_0(\vec q, \omega + \frac{i}{\tau})}{1+ (1- i \omega \tau)^{-1}\left( \frac{\chi_0(\vec q, \omega + \frac{i}{\tau})}{\chi_0(\vec q, 0)}-1 \right)},$
%\begin{equation}
%\chi_M(\vec q, \omega) = \frac{\chi_0(\vec q, \omega + \frac{i}{\tau})}{1+ (1- i \omega \tau)^{-1}\left( \frac{\chi_0(\vec q, \omega + \frac{i}{\tau})}{\chi_0(\vec q, 0)}-1 \right)},
%\label{MerminResponse}
%\end{equation}  
where $\chi_0(\textbf q, \omega)$ is the Lindhard function and $\tau^{-1}$ is the scattering rate. Replacing the polarization above with the Mermin susceptibility, and noting that the Lindhard function for small momentum transfers behaves as $\chi_0(\textbf{q},\omega) = \frac{n q^2}{M\omega^2}$ ($n, M$ are the electron density and mass respectively, with $M$ set to unity henceforth), it is easy to see that the longitudinal conductivity reduces to the Drude formula. The fact that long wavelength optical measurements that are predominantly sensitive to the \textit{transverse} optical conductivity (such as in ellipsometry), also observe \cite{Tompkins2005} a Drude response, points to an important symmetry between the longitudinal and transverse optical conductivities that characterize simple metals, and there is no reason to expect this symmetry to hold in the presence of interactions. To make this symmetry more transparent, we write the finite momentum, $i$-th component of the current-current response of a disordered electron gas minimally coupled to the electromagnetic field as \cite{Altland-Simons}
\begin{eqnarray}\nonumber
 K_i(\bs q_{\perp}, q_{\parallel}, \omega) &=& \frac{1}{4 \pi} \int^{+\infty}_{-\infty} d\nu \, F(\nu, \omega) \, \int d\bs p \,  (2 \bs p + \bs q)^2_i \\
 && \times G^-_{\bs p} (\nu ) G^+_{\bs p + \bs q} (\nu + \omega). 
 \label{Drude}
\end{eqnarray} 
Here the function $F(\nu,\omega)$ contains thermal factors, and the impurity averaged retarded and advanced Green functions are given by $[G^{\pm}_{\bs p}(\nu)]^{-1} = \nu - \xi_{\bs p} \pm \frac{i}{2 \tau}$ where $\xi_{\bs p}$ is the free electronic dispersion measured with respect to the chemical potential. The direction of the momentum transfer $i$ appearing in the electromagnetic vertex is fixed by the driving field, while the momentum transfer $\bs q$ in the Green function contains components both along and perpendicular to the driving field. It is evident from Eq.~\ref{Drude} that one obtains the same value of the conductivity regardless of which component of the momentum transfer ($\bs q_{\perp}$ or $ \bs q_{\parallel}$) is taken to zero first, i.e, $\left[\lim_{q_{\parallel}\rightarrow 0}, \lim_{\textbf{q}_{\perp}\rightarrow 0}\right] K_i(\bs q_{\perp}, \bs q_{\parallel}, \omega) = 0$.  For a fixed direction of the driving field $i$, this statement is true \textit{even in the presence of anisotropy} (in $\xi_{\bs p}$, for example). On the other hand, when $\bs q$ is taken to zero along a fixed direction and the driving field is rotated to be either longitudinal or transverse, as per our expectations, Eq.~\ref{Drude} gives different results only when there is anisotropy. 
\par
\textit{Interactions:} The frequency of the driving field and the scattering rate form two important energy scales in the Drude problem, and together are responsible for the commutation of the two zero momentum limits. Since all energy scales originating from the external field coupling to electrons are compared to these scales, it does not matter along which direction the momentum transfer is taken to zero. This conclusion is robust to the addition of fields with power-law type propagators, as the scales are simply transferred over the new degrees of freedom. One possible way interactions could modify this result is through electron-boson vertices of the form
\begin{equation}
\mathcal{L} = \mathcal{L}_0 + \int d \bs s \, \Theta(\bs p, \bs s)^2 \, \bar{\psi}(\bs p - \bs s) \phi(\bs s) \psi(\bs p) \quad
\label{Lagrangian}
\end{equation} 
where $\Theta(\bs p, \bs s)^2$ is the interaction matrix elements and $\psi(\bs p)$ and $\phi(\bs p)$ are electron and boson fields respectively. To obtain non-commutativity of the two limits, the interaction vertex above cannot originate from electrons interacting with the electromagnetic field or phonons. This state of affairs obtains because the vertex interaction between electrons and light is strictly determined by minimal coupling, and electron-phonon vertices diverge at low momenta (screening of these interactions only introduces an additional scale, and the electron-phonon vertex vanishes in the zero momentum limit) making them unviable. \par
In this work, we envisage electron-boson interactions emerging from electron correlations with matrix elements that depend only on the angles between all the participating momenta. The simplest form of such coupling are functions that depend on the projection of one electron scattered momentum onto the other, i.e $f(\hat{\bs s}.\hat{\bs p})$. The matrix elements cannot be odd in either of the momenta (else the contribution vanishes due to antisymmetry); hence we require vertices at least quartic in the electron momentum. Vertices of such sort originate from electron correlations that depend on gradients of electron fields and have been used frequently to study interaction-driven spontaneously broken rotational symmetries~\cite{Fradkin2001} and coupling between spinless electrons and Goldstone modes~\cite{Vishwanath2014}. In our case, the bosons are obtained as Hubbard-Stratonovich fields through mean field solutions of interactions of the form $\mathcal{L}_{int} = g (\hat {\nabla} \bar{\psi}(\bs r).\hat{\nabla} \psi(\bs r))^2$, where the hat denotes unit vectors along gradients. In other words, we consider interactions where the scattering amplitude depends only on the direction of the scattered momenta and not their magnitude. This motivates a generalized vertex interaction using functions of the form $\Theta(\bs p, \bs q; \bs s)=(\hat{\bs p} + \hat{ \bs q}).(\hat{ \bs p} + \hat{ \bs q}-  \hat{\bs s})$. Using this definition, the matrix elements appearing in Eq.~\ref{Lagrangian} can be written as $\Theta(\bs p, \bs s) \equiv\Theta(\bs p, 0; \bs s)$.  Angular vector interactions of this type are generalizations of effective interactions that have frequently appeared in the literature in the context of dielectric functions on a lattice~\cite{Bagchi1969, Maksimov1981-RMP, Maksimov1978, Keldysh2012-Book}. In these scenarios, angular terms arise due to coupling of long-range Coulomb interactions with either elastic lattice vibrations~\cite{Keldysh2012-Book} or modulations of the Wigner lattice~\cite{Bagchi1969}. However, it must be noted that both the longitudinal and transverse 'modes' studied in these works are defined with respect to polarization of the lattice vibrations. Hence, \text{both} the longitudinal and transverse limits are obtained entirely from the density response, unlike our case where the transverse limit can only be obtained from the current response. \par% \textcolor{blue}{MOTIVATE QUADRUPOLE}\par
\begin{figure}
\centering
\begin{tikzpicture}
\draw (0,0)[line width=0.5mm] ellipse (1cm and 0.5cm);
\draw [decorate,decoration={snake,amplitude=.4mm,segment length=2mm}]
(1,0) -- (1.5,0);
\draw [decorate,decoration={snake,amplitude=.4mm,segment length=2mm}]
(-1,0) -- (-1.5,0);
\node[] at (0,0.8) {(a)};

\draw (4,0) ellipse (1cm and 0.5cm);
\draw [decorate,decoration={snake,amplitude=.4mm,segment length=2mm}]
(5,0) -- (5.5,0);
\draw [decorate,decoration={snake,amplitude=.4mm,segment length=2mm}]
(3,0) -- (2.5,0);
\draw[black,fill=black] (3,0) circle (0.8ex);
\node[] at (4,0.8) {(b)};

\draw (0,-2) ellipse (1cm and 0.5cm);
\draw [decorate,decoration={snake,amplitude=.4mm,segment length=2mm}]
(1,-2) -- (1.5,-2);
\draw [decorate,decoration={snake,amplitude=.4mm,segment length=2mm}]
(-1,-2) -- (-1.5,-2);
\draw [dashed] (0,-2-0.5) -- (0,-2+0.5);
\draw (0,-2-0.5)node[cross]{};
\draw (0,-2+0.5)node[cross]{};
\node[] at (0,-2+0.8) {(c)};

\draw (4,-2) [line width=0.5mm] ellipse (1cm and 0.5cm);
\draw [decorate,decoration={snake,amplitude=.4mm,segment length=2mm}]
(4+1,-2) -- (4+1.5,-2);
\draw [decorate,decoration={snake,amplitude=.4mm,segment length=2mm}]
(4-1,-2) -- (4-1.5,-2);
\draw [dashed] (4,-2-0.5) -- (4,-2+0.5);
\draw (4,-2-0.5)node[cross]{};
\draw (4,-2+0.5)node[cross]{};
\node[] at (4,-2+0.8) {(d)};
\end{tikzpicture}

\caption{Feynman diagrams contributing to the current response in the presence of impurities and electron-boson interactions. The thick (thin) lines denote impurity averaged (free) electron Green functions and the single solid dot denotes vertex corrections from impurities. The dashed line denotes the bosons and the cross symbols are the electron-boson interaction vertices described in the text. } \label{Feynman}
\end{figure}
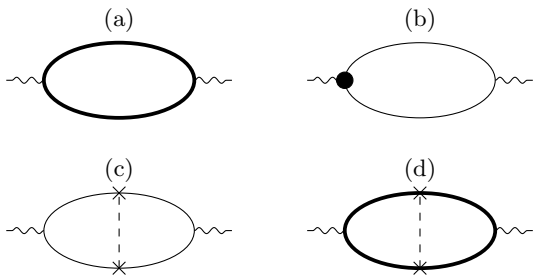
Fig.~\ref{Feynman} shows a set of bubble diagrams that contribute to the current response in the presence of impurities and electron-boson interactions. The Drude formula at finite momentum transfer (appearing in Eq.~\ref{Drude}) is described in Fig.~\ref{Feynman}(a), and gives the response of impurity averaged electrons (thick lines) to an external electromagnetic field. Note that this diagram does not include impurity vertex corrections of the form shown in Fig.~\ref{Feynman}(b) (black solid dot).  That is, the Drude formula neglects impurity lines connecting Green functions across the bubble. This assumption is justified in the limit of zero momentum transfer since the diagram vanishes~\cite{Altland-Simons} in this limit; diagrams of such type will also not be of interest to us as they do not contribute to the non-commutativity. The diagram in Fig.~\ref{Feynman}(d) contains two kinds of perturbative corrections$-$one from impurities and the other from electron-boson interactions$-$and hence can be neglected to lowest order. All that remains now is to evaluate the diagram in Fig.~\ref{Feynman}(c).\par
The lowest order correction to the $i$-th component of the current response, $\delta K_i(\bs q_{\perp}, q_{\parallel}, \omega_q)$, due to the aforementioned interactions is given by
 \begin{eqnarray}
\delta K_i(\bs q_{\perp}, q_{\parallel}, \omega_q)  & = & \int \, d\bs s \, d\bs p \, V^i_{\bs p,\bs q,\bs s} \, \mathcal{M}(\bs s,\bs p; q), \label{Correction}\\ 
\mathcal{M}(\bs s,\bs p; q) &=&\frac1\beta\sum_{i\omega_p}\, \mathcal{G}(p) \, \mathcal{G}(p+q)\, \bar{\mathcal{M}}(\bs s; p, q), \, \nonumber \\ 
\bar{\mathcal{M}}(\bs s; p, q) &=& \frac1\beta \sum_{i\omega_s} \, \mathcal{G}_\phi(s) \, \mathcal{G}(p+q-s) \, \mathcal{G}(p-s) . \nonumber
\label{barM}
 \end{eqnarray}
Here we have used italic variables to denote both momentum and Matsubara frequency, $q=(\bs q, \omega_q)$, and the non-interacting electronic and bosonic (with mass $m$) Green functions are denoted by $\mathcal{G}(p) =  (i \omega_n - \xi_{\bs p})^{-1}$ and $\mathcal{G}_\phi(s) = (\omega_n^2 + \epsilon_\bs s^2)^{-1} $ respectively, with $\epsilon_{\bs s}^2 = m^2 + |\bs s|^2$. The function $V^i_{\bs p,\bs q,\bs s}$ contains all the vertices originating from electrons coupled to the external electromagnetic field and the bosonic field $\phi(s)$, and is given by
\begin{equation}
V^i_{\bs p,\bs q,\bs s} = 4 \left[ (\bs p)_i^2 -  \bs s_i \, (\bs p)_i \right] \,  \Theta(\bs p, \, \bs s)^2 \Theta(\bs p, \bs q; \, \bs s)^2.
\label{Vertices}
\end{equation}
Computing $\delta K_i(\bs q_{\perp}, q_{\parallel}, \omega_q)$ for generic values of frequency and momentum transfer is fairly complicated, especially at non-zero temperatures. Hence we wish to make some simplifying assumptions to maintain analytical tractability. It will be evident later that lifting these assumptions will not qualitatively change the conclusions of our result. To begin, we will assume that the energy cutoff for the boson ($\epsilon_{D}$) is much smaller that the Fermi energy ($\epsilon_F$), and ignore the contributions from fast moving bosons (or large momentum exchanges of the order of the Fermi momentum, $\bs p_F$). This allows us to define a momentum cutoff for the bosons, $\bs s_D$, such that $|\bs s_D| \ll |\bs p_F|$. We will also assume that the frequency of the external field is an intermediate scale in the problem such that $|\bs p_F|^2 \gtrsim \omega \gg |\bs p_F| |\bs s_D|$. Such an assumption is justified since we are interested in energies on the order of the plasma frequency. Using these assumptions and after analytically continuing $i\omega_q$, we simplify the integrand at zero temperature,
\begin{equation}
\mathcal{M}(\bs s,\bs p; q) \simeq \frac{\delta\left( |\bs p| - |\bs p_F| \right)}{ v_F \, \epsilon_{\bs s} \, \omega^2 } \left\{ 1-  \frac{ \left(\bs p \cdot \bs s  \right)^2}{\omega^2}  \right\}.
\label{Integrand}
\end{equation}
In arriving at this form we have dropped contributions from momenta larger than $|\bs p_F|$ since they can be neglected compared to those from momenta on the order of $|\bs p_F|$. Since the momentum transfer $\bs q$ appearing in $\mathcal{M}(\bs s,\bs p; q)$ does not contribute to the discontinuity in the order of limits, we have also set $\bs q$ to zero in the above expression. The Supplemental Material contains details of the calculations.
\par
We can now substitute the integrand appearing in Eq.~\ref{Integrand} back into $\delta K_i(\bs q_{\perp}, q_{\parallel}, \omega_q)$ in Eq.~\ref{Correction}, and perform integrals over the momenta $\bs p$ and $\bs s$. In doing so, odd powers in either $\bs p$ or $\bs s$ vanish due to angular averages, and terms independent of $\hat{\bs q}$ do not contribute to the discontinuity of the limits and can be ignored. Hence the effective vertex function contributing to the discontinuity of the limits is given by $\bar{V}^i_{\bs p, \bs q,\bs s}  =  4 (\hat{\bs q} \cdot \hat{\bs s})^2 \left[ (\bs p)_i^2  + 2 \bs s_i (\bs p)_i (\hat{\bs p} \cdot \hat{\bs s}) + (\bs p)_i^2 (\hat{\bs p} \cdot \hat{\bs s})^2  \right]$. Using this we compute $\delta K_i(\bs q_{\perp}, q_{\parallel}, \omega_q)$ and approach the zero momentum transfer limit in two different ways, the values of which differ by
\begin{eqnarray}
\left[\lim_{q_{\parallel}\rightarrow 0}, \lim_{\textbf{q}_{\perp}\rightarrow 0}\right] K_i(\bs q_{\perp}, q_{\parallel}, \omega) = \frac{\eta}{\omega^2} + \mathscr{O}\left(\frac{1}{\omega^4}\right),
\end{eqnarray}
where $\eta$ is a non-vanishing number which depends on microscopic details of the model.  \par%We demonstrate the computation of $\eta$ for a given direction $i$ in the Supplemental Material. \par
The reason for the discontinuity can be understood through a closer inspection of the factor $(\hat{\bs q}.\hat{\bs s})^2$ appearing in the effective vertex function $\bar{V}^i_{\bs p, \bs q,\bs s}$. This factor can be rewritten and decomposed into longitudinal and transverse components as $(\hat{\bs q}.\hat{\bs s})^2 = \frac{(\bs q. \bs s)^2}{|\bs q|^2|\bs s|^2} = \frac{\left( q_{\parallel}s_{\parallel} + \bs q_{\perp}.\bs s_{\perp} \right)^2}{\left(q_{\parallel}^2 + |\bs q|_{\perp}^2\right)|\bs s|^2}$. It is now evident that this factor picks up the perpendicular (parallel) component of $\bs s$ when $q_{\parallel} (\bs q_{\perp})$ is taken to zero before $\bs q_{\perp} (q_{\parallel})$. Despite the integration over the internal momentum $\bs s$, this feature of the vertex results in different contributions to the current response \textit{even in an isotropic system}. This occurs due to the fact that vertex interactions originating from electron-boson coupling enforce the driving field to preferentially select the direction $i$; by itself, the driving field would have otherwise yielded a commuting order of limits as in the Drude case (see Eq.~\ref{Drude}). On the other hand, if the direction of the driving field was averaged over, one would require \textit{both} vertex interactions and anisotropy (in the band structure, for example) to give non-commutativity of the two limits. Moreover, the discontinuity persists (even for isotropic band structures) irrespective of how the transverse and longitudinal limits are taken to zero, that is, by either keeping the driving field or the momentum transfer fixed. Interestingly, even though there is a discontinuity of the two limits in both isotropic and anisotropic materials, there is a crucial distinction between the two cases. This is evident by observing that the magnitude of the discontinuity is independent of the direction of the driving field only for the isotropic case.  

Since the proposed vertex interaction does not contain any temporal gradients, our work does not address any mismatch in the frequency dependence between the zero momentum limits of the longitudinal and transverse dielectric response. Such temporal fluctuations that are anisotropic can, in principle, give rise to different frequency dependences in a multi-orbital system, and will be the subject of future work. A more quantitative comparison between optical measurements and EELS also demands a better theoretical understanding of the experimental geometries involved, as well as the relationship between bulk response functions and those of the surface. As mentioned in the introduction, corrections to the cross-section resulting from off-normal incidence in ellipsometry has the potential to obscure the interpretation of the data. In reality, the measured dielectric function contains both transverse and longitudinal components, and any experimental adjustment to off-set the longitudinal contribution needs to be considered with caution. Such ambiguity does not arise in EELS  since what is measured is the density-density response function which is directly related to the longitudinal dielectric response function. \par
In conclusion, it is natural to anticipate that the longitudinal and transverse dielectric functions converge to the same value when the momentum transfer is taken to zero (for a fixed direction of the driving field). While this expectation typically holds true in simple metals, there is no reason to believe that it holds in the presence of correlations. In this paper, we proposed a class of electron interactions which gives rise to a discontinuity of the longitudinal and transverse dielectric function in the zero momentum limit. The discontinuity occurs even if the system is inherently isotropic. Since they appear through vertex corrections and depend solely on the relative angles between the momenta of the scattered electrons, the applied external field has the effect of picking a direction in real space but does not add any additional scale to the interactions. This feature of the interaction distinguishes whether the wave vector is taken to zero along the driving field or perpendicular to it. We demonstrated this claim for the case of Drude conductivity modified by the proposed interaction vertices, and summarized consequences of the result to optical and EELS data. This is the first demonstration of how these limits fail to commute at zero momentum.\par
\textit{Acknowledgments:} We thank A.J. Leggett for pointing out Ref.~\cite{Leggett2016} and discussing the subtleties involving experimental measurement of longitudinal and transverse dielectric functions. C.S. and P.W.P. acknowledge support from Center for Emergent Superconductivity, a DOE Energy Frontier Research Center, Grant No. DE-AC0298CH1088.  We also thank the NSF DMR-1461952 for partial funding of this project. K.L. is supported by the Department of Physics at the University of Illinois and a scholarship from the Ministry of Science and Technology, Royal Thai Government. P.A. acknowledges support from the EPiQS program of the Gordon and Betty Moore Foundation, grant GBMF4542. M.M. acknowledges support by the Alexander von Humboldt Foundation through the Feodor Lynen Fellowship program. 
\bibliographystyle{apsrev4-1}%Choose a bibliograhpic style
\bibliography{ZeroMomentum.bib}

\onecolumngrid
\section*{SUPPLEMENTAL MATERIAL}
Here we simplify the correction to the $i$-th component of the current response $\delta K_i(\bs q_{\perp}, q_{\parallel}, \omega_q)$ appearing in Eq. 4 and obtain the expression in Eq. 6.%\ref{Integrand}
 We begin with $\bar{\mathcal{M}}(\bs s; p , q)$,
%\ref{barM }
which is a Matsubara sum over  bosonic frequencies $\omega_s$, 
\begin{eqnarray}
&& \bar{\mathcal{M}}(\bs s; p , q) \equiv \frac1\beta \sum_{i\omega_s} \, \mathcal{G}_\phi(s) \, \mathcal{G}(p+q-s) \, \mathcal{G}(p-s) 
\nonumber \\ && \xrightarrow{i \omega_s \rightarrow z} 
 \oint \frac{dz}{2 \pi i}  \,n_B(z)\, \frac{-1}{z^2-\epsilon_{\bs s}^2} \, \frac{1}{z - (i\omega_p - \xi_{\bs p - \bs s})} \, \frac{1}{z - (i \omega_{p+q} - \xi_{ \bs p+ \bs q- \bs s})} 
\nonumber \\ && = - \sum_{z_i}^4 \text{Res}(z_i) \, n_B(z_i) =  \sum_i^4\bar{\mathcal{M}}^{(i)} \, .
\end{eqnarray}
The sum over $i$ accounts for the contribution coming from the four poles (two bosonic and two fermionic). We use $\omega_{p+q}$ as a shorthand for $\omega_p + \omega_q$. Since $n_{B, F}(i \omega_n + z) = n_{B,F}(z)$ $\left(- n_{F,B}(z)\right)$ when $\omega_n$ is bosonic (fermionic), we have
\begin{subequations}
\begin{align}
\bar{\mathcal{M}}^{(1)} &= \frac{n_B(\epsilon_{\bs s})}{2\epsilon_{\bs s}(i \omega_p - \epsilon_{\bs s} - \xi_{\bs p - \bs s} )( i \omega_{p+q} - \epsilon_{\bs s} - \xi_{\bs p + \bs q - \bs s} )} \, ,
\\ 
\bar{\mathcal{M}}^{(2)} &= \frac{-n_B(-\epsilon_{\bs s})}{2\epsilon_{\bs s}(i \omega_p + \epsilon_{\bs s} - \xi_{\bs p - \bs s} )( i \omega_{p+q} + \epsilon_{\bs s} - \xi_{\bs p + \bs q - \bs s} )} \, ,
\\
\bar{\mathcal{M}}^{(3)} &= \frac{n_F(-\xi_{\bs p - \bs s})}{\left( (i \omega_p - \xi_{\bs p - \bs s})^2 - \epsilon_{\bs s}^2 \right) (i \omega_q + \xi_{\bs p - \bs s} - \xi_{\bs p + \bs q - \bs s})} \, ,
\\
\bar{\mathcal{M}}^{(4)} &= \frac{-n_F(-\xi_{\bs p + \bs q - \bs s})}{\left( (i \omega_{p+q} - \xi_{\bs p + \bs q - \bs s})^2 - \epsilon_{\bs s}^2 \right) (i \omega_q + \xi_{\bs p - \bs s} - \xi_{\bs p + \bs q - \bs s})} \, .
\end{align}
\end{subequations}
In order to obtain $\mathcal{M}(\bs s, \bs p; q)$, we now perform a Matsubara sum over the fermionic frequency $\omega_p$,
\begin{eqnarray}
\mathcal{M}^{(i)} (\bs s, \bs p; q) = \frac1\beta \sum_{i \omega_p} \, \mathcal{G}(p) \, \mathcal{G}(p+q)\, \bar{\mathcal{M}}^{(i)} (\bs s; p, q) .
\end{eqnarray}
Like before, we have defined $\mathcal{M} = \sum_i^4 \mathcal{M}^{(i)} $. Since eventually we are interested in the limit of $|\bs q| \rightarrow 0$, we evaluate the above expression setting  to order $\mathscr{O}(|\bs q|^0)$. This is done by assuming $v_F |\bs q| \ll \omega_q$, where $v_F$ is the Fermi velocity of the electrons. Due to this assumption, our final expression may not be used to recover the conductivity in the DC limit, $\omega \rightarrow 0$. Denoting $\xi_{\bs p,\bs s}^\pm = \xi_{\bs p - \bs s} \pm \epsilon_{\bs s}$ and using the previous identities we simplify the residues as 
\begin{subequations}
\begin{align}
\mathcal{M}^{(1)} &= - \frac{n_B(\epsilon_{\bs s})}{\epsilon_{\bs s}} \left\{ 
\frac{n_F(\xi_{\bs p}) - n_F(\xi^+_{\bs p, \bs s})}{(\xi^+_{\bs p, \bs s} - \xi_{\bs p})\left[ \omega_q^2 + (\xi^+_{\bs p, \bs s} - \xi_{\bs p})^2 \right] }  \right\} \, ,
\\
\mathcal{M}^{(2)} &= \frac{n_B(- \epsilon_{\bs s})}{\epsilon_{\bs s}} \left\{ 
\frac{n_F(\xi_{\bs p}) - n_F(\xi^-_{\bs p, \bs s})}{(\xi^-_{\bs p, \bs s} - \xi_{\bs p})\left[ \omega_q^2 + (\xi^-_{\bs p, \bs s} - \xi_{\bs p})^2 \right] }  \right\} \, ,
\\
\mathcal{M}^{(3)} &= - \frac{n_F(-\xi_{\bs p - \bs s})}{i \omega_q} \Bigg\{ 
\frac{ n_F({\xi_{\bs p})}(i \omega_q - 2\xi_{\bs p} + 2 \xi_{\bs p - \bs s}) }{[(\xi_{\bs p} - \xi_{\bs p - \bs s})^2 - \epsilon_{\bs s}^2][(i \omega_q - \xi_{\bs p} + \xi_{\bs p - \bs s})^2 - \epsilon_{\bs s}^2]} \nonumber \\ & 
+ \frac{n_F(\xi^+_{\bs p, \bs s})}{2\epsilon_{\bs s}(\xi^+_{\bs p, \bs s}-\xi_{\bs p})(i \omega_q - \xi_{\bs p} + \xi^+_{\bs p, \bs s}) } 
- \frac{n_F(\xi^-_{\bs p, \bs s})}{2\epsilon_{\bs s}(\xi^-_{\bs p, \bs s}-\xi_{\bs p})(i \omega_q - \xi_{\bs p} + \xi^-_{\bs p, \bs s}) } 
\Bigg\}
 \, ,
\\
\mathcal{M}^{(4)} &= \frac{n_F(-\xi_{\bs p - \bs s})}{i \omega_q} \Bigg\{ 
\frac{n_F({\xi_{\bs p})}( i \omega_q - 2\xi_{\bs p} + 2 \xi_{\bs p - \bs s}) }{ \left[(\xi_{\bs p} - \xi_{\bs p - \bs s})^2 - \epsilon_{\bs s}^2 \right] \left[(i \omega_q + \xi_{\bs p} - \xi_{\bs p - \bs s})^2 - \epsilon_{\bs s}^2 \right]} \nonumber \\ & 
+ \frac{n_F(\xi^+_{\bs p, \bs s})}{2\epsilon_{\bs s}(\xi^+_{\bs p, \bs s}-\xi_{\bs p})(i \omega_q + \xi_{\bs p} - \xi^+_{\bs p, \bs s}) } 
- \frac{n_F(\xi^-_{\bs p, \bs s})}{2\epsilon_{\bs s}(\xi^-_{\bs p, \bs s}-\xi_{\bs p})(i \omega_q + \xi_{\bs p} - \xi^-_{\bs p, \bs s}) } 
\Bigg\}
 \, ,
\end{align}
\end{subequations}
where all the thermal sums have been performed. An immediate consequence of our scale choice $p_F^2 \gg s_D^2$ is that $ n_F(\xi_{\bs p}) - n_F(\xi^+_{\bs p, \bs s}) \approx 0$, meaning the low lying electrons do not contribute significantly to the scattering processes and it is coming from around $\bs p \simeq \bs p_F$. This can be achieved by writing, $\left[n_F(\xi^\pm_{\bs p, \bs s}) - n_F(\xi_{\bs p}) \right]/\left[ \xi^\pm_{\bs p, \bs s} - \xi_{\bs p} \right] \simeq \partial \, n_F(\xi_{\bs p}) \, \simeq -  \delta(\xi_{\bs p}) = - \delta(|\bs p| - p_F)\,/v_F$. These equalities are exact at zero temperature. Using this we can simplify $\mathcal{M}^{(1, 2)}$ to
\beq
- \frac{\mathcal{M}^{(1)} }{n_B(\epsilon_{\bs s})} = \frac{\delta(|\bs p| - p_F)}{v_F \, \epsilon_{\bs s} \left\{ \omega_q^2 + (\xi_{\bs p} - \xi_{\bs p, \bs s}^-)^2 \right\} }  = \frac{\mathcal{M}^{(2)} }{n_B(-\epsilon_{\bs s})} \, .
\eeq
Since at zero temperature $n_{F,B}(x) = \pm 1, \forall x < 0$ and zero otherwise, $n_B(\epsilon_{\bs s}) = 0$ making $\mathcal{M}^{(1)} =0$ and $n_B(-\epsilon_{\bs s})=-1$. The fermionic functions can also be simplified as 
\begin{eqnarray}
n_F(\xi^\pm_{\bs p, \bs s}) \approx \begin{cases}
n_F(- \mu) = 1&, |\bs p| \in (0, p_F) \\
n_F( \pm m) = 0 \,\text{or}\, 1& , |\bs p| \in (p_F, \infty) 
\end{cases}
\,\, \text{and} \,\,\,
n_{F}(\pm \xi_{\bs p}) = \begin{cases}
1 \,\,\text{or}\,\, 0 &, |\bs p| \in (0, p_F) \\
0 \,\,\text{or}\,\, 1 &,  |\bs p| \in (p_F, \infty) 
\end{cases} \, .
\nonumber \\
\label{eq:NBFsimple}
\end{eqnarray}
The above equations imply the first terms in $\mathcal{M}^{(3,4)}$ are zero for all $|\bs p|$ values. The rest of the terms can be added and simplified to [using Heaviside step function $H(x)$]
\begin{eqnarray}
\mathcal{M}^{(3)} + \mathcal{M}^{(4)} &=& \frac{- n_F(-\xi_{\bs p - \bs s})}{i \omega_q \epsilon_{\bs s}} \left\{ \frac{n_F(\xi^+_{\bs p, \bs s})}{\omega_q^2 + (\xi_{\bs p} - \xi_{\bs p, \bs s}^+)^2} - \frac{n_F(\xi^-_{\bs p, \bs s})}{\omega_q^2 + (\xi_{\bs p} - \xi_{\bs p, \bs s}^-)^2}
\right\} \, 
\nonumber \\ &=& 
H( |\bs p| - p_F) \left\{i \omega_q \epsilon_{\bs s}  \left( \omega_q^2 + (\xi_{\bs p} - \xi_{\bs p,\bs s}^-)^2 \right) \right \}^{-1}
\end{eqnarray}
Combining all the above manipulations we have the final expression of $\mathcal{M}(\bs p, \bs s; q)$ at zero temperature,
\beq
\mathcal{M}(\bs p, \bs s; q)  = \sum_i \mathcal{M}^{(i)} = 
\left[ \frac{H(|\bs p| - | \bs p_F|)}{i \omega_q \epsilon_{\bs s}} - \frac{\delta(|\bs p| - |\bs p_F|)}{v_F \, \epsilon_{\bs s}} \right] \, \left\{ \omega_q^2 + (\xi_{\bs p} - \xi_{\bs p, \bs s}^-)^2 \right\}^{-1}.
\eeq
Our assumption of $\omega_q$ being an intermediate scale along with other scale choices can be used to further simplify the above expression. Since the contributions from momenta much larger than the Fermi momentum is much smaller than those around the Fermi level, we ignore the Heaviside term and simplify the above equation using
\begin{eqnarray}
\left. \left\{ \omega_q^2 + (\xi_{\bs p} - \xi_{\bs p,\bs s}^-)^2 \right\}^{-1} \right\vert_{\bs p\, = \, \bs p_F} &=& \left\{ \omega_q^2 + \left( \epsilon_{\bs s} - \frac{|\bs s|^2}{2} - {\bs p_F \cdot \bs s} \right)^2 \right\}^{-1} 
\nonumber \\ & \simeq &
\frac{1}{\omega_q^2 } \, \left\{ 1 + \left( \frac{\bs p_F \cdot \bs s}{ \omega_q} \right)^2 \right\}^{-1} \quad \left( |\bs p_F| |\bs  s_D| \gg s_D^2 \right)
\nonumber \\ & \simeq &
\frac{1}{\omega_q^2 } \, \left\{ 1 - \left( \frac{\bs p_F \cdot \bs s}{ \omega_q} \right)^2 \right\} \qquad \left( |\bs p_F| |\bs  s_D| \ll \omega_q \right) \, .
\end{eqnarray}
Plugging the simplification back into $\mathcal{M}(\bs p, \bs s; q)$, we obtain the expression Eq. 6 %\ref{Integrand}
 in the main text.

\end{document}